\documentclass[prb,twocolumn,showpacs,amsmath,letter]{revtex4}
\usepackage{graphicx}

\newcommand{\Journal}[4]{#1 \textbf{#2}, #3 (#4)}

\begin{document}

\title{Spin-Transfer Torque and Electron-Magnon Scattering}

\author{S. Urazhdin}
\affiliation{Department of Physics and Astronomy, Johns Hopkins
University, Baltimore, MD, 21218}


\begin{abstract}

According to the spin-torque model, current-driven magnetic dynamics in ferromagnetic multilayers is determined by the transfer of electron spin perpendicular to the layers' magnetizations. By separating the largest contributions to the magnetic dynamics, we demonstrate that the dominant effect of spin-torque is rather due to the electron spin parallel to the field. We show that this effect can be equivalently described as stimulated current-driven excitation of spin-waves, and discuss four specifically quantum-mechanical aspects of spin-transfer, not described by the spin-torque.

\end{abstract}
\maketitle

Current-induced magnetization dynamics has recently received significant attention due to the potential applications in the magnetic memory devices, generation of microwaves, and field sensors. Studies of this effect have also offered new insights into the fundamental properties of exchange interaction between the conduction electrons and the magnetization in ferromagnets. According to the generally accepted spin-transfer torque model (STT), current-induced magnetic excitation and magnetization switching are caused by the absorption of the spin current transverse to the magnetization. Spin conservation arguments are then used to determine the current-driven magnetic dynamics, which is usually described by the Landau-Lifshitz equation with an additional current-dependent STT term~\cite{slonczewski,suntheory}.

Predictions of STT have been qualitatively, and in some cases quantitatively supported by experiments~\cite{cornellorig,koch}. However, a clear connection with the fundamental formalism of spin-wave (magnon) interactions with electrons has not yet been established. Moreover, STT is sometimes described as a new interaction mechanism~\cite{ralphscience,koch}, and the relative roles of STT-induced precession {\it vs.} incoherent spin-wave generation by electron-magnon scattering are being debated~\cite{bazaliy,mytheory,cornelltemp2}. Many experimental and theoretical studies are successfully interpreted with the simplest uniform (macrospin) current-driven precession of the magnetization~\cite{cornellorig,koch,cornelltemp2}. On the other hand, there is indirect experimental evidence~\cite{myprl,wegroweprl}, partly supported by micromagnetic simulations~\cite{micromagnetics}, for the large role of inhomogeneous current-driven magnetic excitations. Those behaviors are most naturally described in momentum and energy representation of spin waves, rather than coordinate space. However, the applications of the STT model are presently essentially limited to the Landau-Lifshitz equation.

The goal of this publication is to establish the connection between STT and the physics of electron-magnon scattering, giving a broader perspective and understanding of the current-driven phenomena. Our results let one avoid the complex calculations of transverse spin currents presently involved in STT analysis; we show that only the transfer of spin parallel to the field is usually important, greatly simplifying the analysis of STT. We also discuss specifically quantum-mechanical aspects of spin-transfer, not described by STT approximation.

{\it Spin-Transfer Torque.} In the spin-torque model, all the electrons flowing towards a ferromagnetic layer F contribute equally to the spin-transfer, regardless whether they are reflected or transmitted by F, i.e. they may or may not contribute to the charge current~\cite{slonczewski,stiles}. For the uniform (macrospin) magnetic dynamics, it is sufficient to simply keep track of the number and spin orientations of electrons scattered by the ferromagnet. In a macrospin approximation, we can reduce STT to a model involving the scattering of two spins, a large ${\mathbf S}$ representing the magnetic moments of F, and a spin $|{\mathbf s}|=1/2$ representing a conduction electron, omitting the spatial coordinates of both~\cite{mytheory}.

The interaction between a conduction electron and the ferromagnet is approximately described by the Stoner exchange potential $E_{ex}=-J{\mathbf s}{\mathbf S}/S$, which gives a rapid precession of ${\mathbf s}$ around ${\mathbf S}$. Here $J=0.1-1$~eV for the transition metal ferromagnets, $S=|{\mathbf S}|$. If $J/\hbar$ is comparable to the inverse of the time an electron spends in the ferromagnet, the particular value of $J$ is unimportant: the role of the exchange interaction is to randomize the spin precession phases of the scattered electrons, due to the strong dependence of those phases on the electron wave vector. Since the precession phase gives the direction of electron spin component perpendicular to the magnetization, all of the initial electron spin transverse to the magnetization is, on average, absorbed by it~\cite{slonczewski,stiles}. For electrons flowing towards the ferromagnet at a rate $I_n=dn/dt$, spin-conservation then gives a current-driven STT term in the Landau-Lifshitz equation for the magnetic dynamics~\cite{slonczewski,suntheory}
\begin{equation}\label{spintorque}
\frac{d{\mathbf S}}{dt}=\gamma {\mathbf S}\times({\mathbf H_{eff}}-\alpha\frac{\mathbf S}{S}\times {\mathbf H_{eff}}+
\tau {\mathbf s}\times\frac{\mathbf S}{S}).
\end{equation}
Here, we for simplicity assume that the anisotropy of F is an ellipsoid of rotation, described by the anisotropy field ${\mathbf H}_{a}$ along the applied field ${\mathbf H}$, and ${\mathbf H}_{eff}={\mathbf H}+{\mathbf H_a}$. $\alpha$ is the Gilbert damping parameter, $\gamma$ is the gyromagnetic ratio, $\tau=I_n/(\gamma S)$. Analysis of Eq.~\ref{spintorque} showed that the current-driven STT term may suppresses Gilbert damping, inducing dynamical instability~\cite{slonczewski,suntheory}.  The crucial new step in our analysis is to separate the dominant terms from the small corrections in Eq.~\ref{spintorque} close to this instability, leading to dependencies quite different from the naive picture of STT. 

Typically, $\alpha\approx 0.01$. The dynamical instability leading to switching occurs when the last two terms in Eq.~\ref{spintorque} are similar~\cite{slonczewski}, and two orders of magnitude smaller than the first term. If they are neglected, Eq.~\ref{spintorque} gives a periodic precession with time $t$ of ${\mathbf S}$ around ${\mathbf H_{eff}}$: $\theta=\theta_0$, $\phi=-\cos(\gamma H_{eff}t)$. We now evaluate the corrections due to damping and STT. For simplicity, we separately consider the cases ${\mathbf s}\parallel{\mathbf H_{eff}}$ and ${\mathbf s}\perp{\mathbf H_{eff}}$.

In the first case, we define ${\mathbf S}=(S_x,S_y,S_z)=(S\sin\theta\cos\phi,S\sin\theta\sin\phi,S\cos\theta)$, ${\mathbf s}=(0,0,s_z)$ for $z$-axis set along ${\mathbf H_{eff}}$. Eq.~\ref{spintorque} then gives
\begin{equation}\label{ssz}
d\phi/dt=\gamma H_{eff};\ d\theta/dt=-\gamma(\alpha H_{eff}+s_z\tau)\sin\theta.
\end{equation}
The first equation describes precession of ${\mathbf S}$ around ${\mathbf H_{eff}}$, the second describes the evolution of the precession amplitude. Eq.~\ref{ssz} shows that STT enhances or suppresses the damping, depending on the sign of $s_z$.

For ${\mathbf s}\perp{\mathbf H_{eff}}$, the effect of STT is quite different. In this case, we replace ${\mathbf H'}={\mathbf H}_{eff}+\tau \lambda {\mathbf s}\times {\mathbf H_{eff}}/|{\mathbf H}_{eff}|$ in Eq.~\ref{spintorque}. The parameter $\lambda\le1$ is determined below. Keeping only terms of up to the first order in $\alpha$, $\tau/H_{eff}$, \begin{equation}\label{spintorque2}
\frac{d{\mathbf S}}{dt}=\gamma {\mathbf S}\times \left[{\mathbf H'}-\alpha\frac{\mathbf S}{S}\times {\mathbf H'}+
\tau {\mathbf s}\times\left(\frac{\mathbf S}{S}-\lambda\frac{\mathbf H'}{H'}\right)\right].
\end{equation}
Eq.~\ref{spintorque2} describes precession of ${\mathbf S}$ around an effective field ${\mathbf H'}$ with a modified STT term.

Similarly to the case of ${\mathbf s}\parallel{\mathbf H_{eff}}$, we define 
${\mathbf S}=(S\sin\theta\cos\phi,S\sin\theta\sin\phi,S\cos\theta)$ in the frame set by ${\mathbf H'}$, and ${\mathbf s}=(s_x,0,0)$ to the first order in $\tau/H_{eff}$ in Eq.~\ref{spintorque2}. By setting $\lambda=\cos\theta_0$, where $\theta(t)=\theta_0+\theta_1(t)$ for small $t$, we eliminate $S_z$ from the last term in Eq.~\ref{spintorque2}, giving
\begin{equation}\label{ssx}
d\phi/dt=\gamma(H'+\tau\sin\theta\sin\phi);\ d\theta/dt=-\gamma\alpha H'\sin\theta.
\end{equation}
Eqs.~\ref{ssx} show that ${\mathbf s}\perp{\mathbf H_{eff}}$ only distorts the magnetization precession trajectory, but does not contribute to the precession damping. To the first order in $\alpha$, $\tau/H_{eff}$, the dynamics described by Eqs.~\ref{ssz},~\ref{ssx} depends linearly on ${\mathbf s}$. Thus, for an arbitrary orientation of ${\mathbf s}$, these equations give the contributions to the magnetic dynamics from the spin components parallel and perpendicular to the field electron, respectively.

Eqs.~\ref{ssz},~\ref{ssx} are the central result of our analysis of STT. We showed that STT can be separated into two contributions: i) enhancement or suppression of magnetization precession damping, determined by the conduction electron spin component along the field, and ii) tilting of the precession axis, determined by the electron spin component perpendicular to the field. Eq.~\ref{ssx} also predicts a weak unharmonic distortion of precession. Since $\tau/H_{eff}\approx 0.01$ near the stability threshold, ${\mathbf H'}$ generally deviates from ${\mathbf H_{eff}}$ by less than $1^\circ$. Thus, the difference between the electron spin components referenced either to ${\mathbf H_{eff}}$ or to ${\mathbf H'}$ is small. 

Eq.~\ref{ssz} shows that the dynamical instability is driven by the electron spin  component along ${\mathbf H_{eff}}$. Moreover, by integrating the STT term in Eq.~\ref{spintorque} over the precession trajectory given by Eq.~\ref{ssz}, we obtain for the spin-transfer along the x-axis
\begin{eqnarray}\label{deltaSx}
\Delta S_x(t)=\frac{\tau Ss_z}{2}\nonumber\\
\frac{H_{eff}\cos2\theta\cos\phi+2\sin2\theta\sin\phi(\alpha H_{eff}+s_z\tau)}{H^2_{eff}-4(\alpha H_{eff}+ s_z\tau)^2}.
\end{eqnarray}
It is a limited oscillating function of time. In contrast, the $z$-component of spin-transfer
\begin{equation}\label{deltaSz}
\Delta S_z(t)=\gamma\tau Ss_z\{t+\sin2\theta/[2\gamma(\alpha H_{eff}+s_z\tau)]\}/2
\end{equation}
is approximately linear with time. Thus, not only the $s_z$ component of electron spin dominates the magnetic dynamics, but also only the $z$-component of spin-transfer is significant, when integrated over the precession trajectory. These results are consistent with the energy conservation arguments of Ref.~7: spin-transfer perpendicular to the field does not affect the Zeeman energy of the ferromagnet, and thus cannot contribute to the precession damping, as does spin-transfer along the field.

Our analysis resolves the controversy regarding the relative importance of spin-currents perpendicular to the magnetization~\cite{slonczewski,suntheory,stiles,polianski,bazaliyrc} or parallel to it~\cite{wegrowejap,mytheory,tsoiprl}: the onset of instability is determined by the component of spin-current, parallel to the {\it equilibrium} magnetization orientation. However, the instability can be equivalently explained by the spin-current perpendicular to the {\it instantaneous} orientation of the precessing magnetization.

Eqs.~\ref{ssz},~\ref{ssx} have important consequences both for modeling the current-driven magnetic dynamics and the fundamental understanding of spin-transfer. Relatively straightforward two spin-channel models of magnetotransport have been extensively developed and used in the studies of the giant magnetoresistance~\cite{valetfert}. In contrast, the nature of STT generally requires that spin-currents and spin-accumulation both parallel and perpendicular to the magnetizations of the ferromagnets are considered, making the analysis significantly more complicated. Eq.~\ref{ssz} demonstrates that one does not need to calculate the transverse spin-currents to determine the onset of current-driven instability and switching. It is sufficient to find the difference between the spin-up and spin-down electron fluxes towards the ferromagnet, as defined in the frame set by the ${\mathbf H_{eff}}$. To illustrate this result, we give a simple analysis of the current-switching experiments with noncollinear magnetizations in trilayers F$_1$/N/F$_2$, where F$_1$ is a fixed (e.g. thick and extended) magnetic layer, $N$ is a nonmagnetic spacer, and F$_2$ is a nanopatterned ferromagnet being switched by the current~\cite{mancoff,iswvsmr}. Eq.~\ref{spintorque} predicts that STT increases with the static angle $\theta$ between magnetizations M$_1$ and M$_2$ of F$_1$ and F$_2$, giving the orientations of ${\mathbf s}$ and ${\mathbf S}$ in our model. However, $I_s\propto 1/\cos\theta$ was experimentally established for the magnetization switching current $I_s$, i.e $I_s$ increases with $\theta$. This dependence directly follows from Eq.~\ref{ssz}: since $s_z=(\cos\theta)/2$, the dynamical instability/switching occurs when STT compensates the damping, giving $I_s\propto 1/\cos\theta$.

{\it Electron-Magnon Scattering.} To establish the  connection between STT and electron-magnon scattering, we introduce the number of magnons $n=S(1-\cos\theta)\approx S\theta^2/2$ for small $\theta$. Eq.~\ref{ssz} then gives
\begin{equation}\label{magnons}
dn/dt\approx-2n(\alpha\gamma H_{eff}+ s_zI_n/S)
\end{equation}
The second term in Eq.~\ref{magnons} describes emission/absorption of magnons by the conduction electrons, at a rate proportional to the magnon density $n/S$. Except for the lack of the specifically quantum-mechanical spontaneous emission term, it is identical to the Einstein formula for electron-magnon scattering~\cite{mytheory}. Thus, the component of STT driving the dynamical instability does not represent a new form of interaction between the conduction electrons and the magnetization. Instead, STT provides a convenient way for calculating the scattering rate between the conduction electrons and magnons in the limit of strong exchange coupling. Does one of these formalisms offer insights the other one does not? On the one hand, the current-driven effective field (Eq.~\ref{spintorque2}) is a specific property of STT model. On the other, the quantum-mechanical description of electron-magnon scattering must be more general, i.e. apply to cases when the classical STT fails. We discuss several such possibilities below. But first, we address a mostly formal distinction in the description of inhomogeneous states.

{\it Inhomogeneous excitations.} Electron-magnon scattering is naturally formulated in the momentum and energy space, while the STT is formulated in the real space for the magnetization coordinates, and does not {\it a'priori} include the energy conservation. In the original works on STT, the energy of the uniform precession was assumed small~\cite{berger} or neglected altogether~\cite{slonczewski}, while the energies of inhomogeneous excitations (spin-waves) were assumed prohibitively large, so that their generation by the current was not energetically allowed. These assumptions are not valid in typical experiments with nanopillars. Consider for example a typical nanopatterned Co layer with $d\approx 100$~nm lateral dimensions. The exchange contribution to the spin-wave energy with wavelength $\lambda=200$~nm is $E_{ex}(\lambda)=2\pi Ag\mu_B/(M\lambda^2)\approx 1$~$\mu$eV. Here $A=10^{-11}$~J/m is the exchange stiffness, $M=1440$~emu/cm$^3$ is the magnetization, $\mu_B$ is the Bohr constant, $g$ is the gyromagnetic ratio. For $H_{eff}\approx1$~kOe, the dipolar energy of the uniform precession mode is similar, $E_d=g\mu_BH_{eff}=10$~$\mu$eV. Since these energies are comparable, and significantly smaller than the energies of the conduction electrons $\approx1$~meV at the typical switching current densities $\approx 10^{7}$~A/cm$^{2}$, current-driven excitation of many spin-wave modes is energetically allowed.

Excitation of inhomogeneous magnetic states by the current can be described by a local STT approximation, i.e. Eq.~\ref{spintorque} extended to a micromagnetic Landau-Lifshitz equation~\cite{micromagnetics}. Complicated current-driven micromagnetic dynamics is usually predicted. An obvious drawback of this formalism is the lack of generality and predictive power, e.g. contained in Eqs.~\ref{ssz},~\ref{ssx} for the macrospin. This model also does not capture the cut-off of the high-energy spin-wave generation (see {\it quantum threshold} below)~\cite{berger,tsoiprl}.

For a particular spin-wave mode with a long wavelength $\lambda$, it is easy to see that Eq.~\ref{magnons} holds independently of $\lambda$: the spin-conservation arguments used to obtain the STT term in Eq.~\ref{spintorque} hold locally for the volume of ferromagent smaller than $\lambda^3$. Only $H_{eff}$ and $\alpha$ depend on $\lambda$, due to the interactions with the rest of the feromagnet. Eq.~\ref{ssz} then also holds locally~\cite{micromagnetics,polianski}. The magnon density can be defined by the local variation of magnetization $n/S=1-\cos\theta$, where $\theta$ is the angle of the local deviation from equilibrium. Thus, Eq.~\ref{magnons} also describes the population dynamics of finite-wavelength spinwaves. Instead of the local version of Eq.~\ref{spintorque}, Eq.~\ref{magnons} can be replaced with a set of equations for individual spin-wave modes $i$ with populations $n_i$, coupled through the nonlinear damping $\alpha_i(n_1,n_2,...)$. This approach naturally predicts highly nonuniform small amplitude current-driven dynamics, justifying the effective magnetic temperature approximation~\cite{mytheory}. However, due to the complicated nonlinear damping $\alpha_i(n_1,n_2,...)$, the real space micromagnetic models may be more attractive for the description of large-amplitude dynamics.

We now consider the quantum-mechanical aspects of spin-transfer, beyond the classical approximation for the current-driven magnetic dynamics used in STT.

{\it Spontaneous magnon emission.} A consistent quantum-mechanical derivation of electron-magnon scattering gives both stimulated (Eq.~\ref{magnons}) and spontaneous current-driven magnon emission~\cite{mytheory}. Spontaneous emission cannot be captured within the classical description of the magnetization in STT. However, at realistic experimental temperatures and applied fields, the thermal populations of a large number of spin-wave modes are large, $n\gg 1$. The spontaneous contribution is then negligible.

{\it The Stern-Gerlach experiment.} Consider a spin-transfer experiment, in which electrons spin-polarized along the applied field ($s_z=\pm1/2$) are scattered by an excited ferromagnet characterized by the number of magnons $n$. After the scattering, the $z$-projections of the electron spins are measured, as in the Stern-Gerlach experiment~\cite{weber}. In the quantum-mechanical language, such a measurement collapses the entangled state of the electrons and the magnetization. It is easy to see that the spin-transfer is $\Delta s_z=0$ or $-1$, $\Delta s_x=\Delta s_y=0$. This result is quite different from the STT arguments based on spin-conservation. The reason for such a difference is that the hamiltonian of the system conserves only the $z$-component of angular momentum, but does not conserve $z$- and $y$-components, resulting in a non-classical magnetization state after the scattering. By calculating the scattering matrix elements, or by applying the correspondence principle to Eq.~\ref{magnons}, one obtains $P(-1/2)=n/S$, $P(1/2)=1-P(-1/2)$ for the probabilities for the final electron spin projections $s_z=-1/2$ and $1/2$, correspondingly. Thus, Eq.~\ref{magnons} for the current-driven dynamics holds for a statistically large number of scattered electrons. However, scattering of individual electrons is not governed by the spin-conservation arguments of STT. This experiment illustrates that a {\it physical difference} between the descriptions of spin-transfer in terms of electron spin-flipping or classical precession averaging (giving STT) appears when the quantum state of the electron after scattering is specified.

{\it Electron-magnon scattering on impurities.} Because of the large mismatch between the wave vectors of the conduction electrons and magnons, momentum conservation severely limits the phase space volume available for electron-magnon scattering in the bulk of ferromagnets. Non-conservation of momentum at the interfaces gives a large electron-magnon scattering amplitude. These arguments are an alternative form of the assumptions behind STT~\cite{berger}. Electron scattering on long-wavelength spin waves in the bulk of ferromagnets has been described in the framework of STT~\cite{bazaliyrc}. However, a more general quantum mechanical approach may be necessary for impurity-mediated electron-magnon scattering~\cite{fert}. The amplitude of electron scattering on a small localized spin ${\mathbf S}$ cannot be described by STT, derived for $S\gg 1$; the quantum aspects of its dynamics become important~\cite{mytheory}.

{\it Quantum threshold for excitations.} Some models of spin-transfer have stressed the importance of the spin-wave excitation gap, setting the threshold electron energy necessary to excite a magnon~\cite{berger,tsoiprl}. At typical current densities $\approx10^7$A/cm$^2$ in spin-transfer experiments, the energies of electrons transversing a thin ferromagnetic layer are $\approx 1$~meV at temperature $T=0$. At realistic experimental temperatures $10-300$~K, the thermal electron energies are $\approx 1-30$~meV. Since the magnon excitation gap is $\Delta\approx 10$~$\mu$eV at $H_{eff}=1$~kOe, we conclude that the quantum threshold is usually completely smeared out. However, at large $H>10$~T, $\Delta>1$~meV may become important at $T<10$~K. The low-energy electrons then cannot excite magnons because of the energy conservation. The spin-conservation arguments of STT must still hold, implying that the precessing electron momentum is transmitted directly to the field, without exciting the ferromagnet. A similar effect may be occur due to a large magneto-crystalline anisotropy~\cite{nunez}. In this case, angular momentum should be transferred directly to the lattice, similar to the Mossbauer effect.

In summary, we have shown that the spin-transfer torque (STT) can be separated into two contributions: i) enhancement or suppression of damping, driven by the spin current component along the effective magnetic field, and ii) an effective field, driven by the  spin component orthogonal to the field. We show that the former can be alternatively described as stimulated electron-magnon scattering. We point out the formal differences between these two formalisms in the description of inhomogeneous magnetization states, and several cases in which electron-magnon scattering is not described by the STT approximation: i) Spontaneous magnon emission, ii) Stern-Gerlach experiment, iii) Scattering on impurities, iv) Mossbauer-type spin rotation of electrons, without recoil on the magnetization when the applied field or magnetic anisotropy is large, related to the quantum threshold for excitations. Future experiments will show the relative importance of these effects in real systems.

The author acknowledges helpful discussions with N.O. Birge and M.D. Stiles, and support from the NSF through Grant DMR00-8031.


\begin{thebibliography}{99}
\bibitem{slonczewski} J. Slonczewski, \Journal{J. Magn. Magn. Mater.}{159}{L1}{1996}; ibid. {\bf 247}, 324 (2002).
\bibitem{suntheory} J.Z. Sun, \Journal{Phys. Rev.}{B 62}{570}{2000}.
\bibitem{cornellorig} J.A. Katine, F.J. Albert, R.A. Buhrman, E.B. Myers and D.C. Ralph, \Journal{Phys. Rev. Lett.}{84}{3149}{2000}.
\bibitem{koch} R.H. Koch, J.A. Katine, and J.Z. Sun, \Journal{Phys. Rev. Lett.}{92}{088302}{2004}.
\bibitem{ralphscience} D.C. Ralph, \Journal{Science}{291}{999}{2001}.
\bibitem{bazaliy} Ya.B. Bazaliy and B.A. Jones, \Journal{Physica B - Cond. Mat.}{329}{1290}{2003}.
\bibitem{mytheory} S. Urazhdin, \Journal {Phys. Rev.}{\bf B 69}{134430}{2004}.
\bibitem{cornelltemp2} I.N. Krivorotov, N.C. Emley, A.G.F. Garcia, J.C. Sankey, S.I. Kiselev, D.C. Ralph, and R.A. Buhrman, \Journal{Phys. Rev. Lett.}{93}{166603}{2004}.
\bibitem{myprl} S. Urazhdin, N.O. Birge, W.P. Pratt Jr., and J. Bass, \Journal{Phys. Rev. Lett.}{91}{146803}{2003}.
\bibitem{wegroweprl} A. Fabian, C. Terrier, S.S. Guisan, X. Hoffer, M. Dubey, L. Gravier, J.-P. Ansermet, and J.-E. Wegrowe, \Journal{Phys. Rev. Lett}{91}{257209}{2003}.
\bibitem{micromagnetics} J. Miltat, G. Albuquerque, A. Thiaville, C. Vouille, \Journal{J. Appl. Phys.}{89}{6982}{2001}; X. Zhu, J.G. Zhu, and R.M. White, \Journal{J. Appl. Phys.}{95}{6630}{2004}; K.J. Lee, A. Deac, O. Redon, J.P. Noziers, and B. Dieny, cond-mat/0406628.
\bibitem{stiles} M. D. Stiles and A. Zangwill, \Journal{Phys. Rev.}{B 66}{014407}{2002}.
\bibitem{wegrowejap} J.E. Wegrowe, X. Hoffer, Ph. Guittienne, A. Fabian, L. Gravier, T. Wade, and J.Ph. Ansermet, \Journal{J. Appl. Phys.}{91}{6806}{2002}.
\bibitem{valetfert} T. Valet and A. Fert, \Journal{Phys. Rev.}{B 48}{7099}{1993}.
\bibitem{mancoff} F.B. Mancoff, R.W. Dave, N.D. Rizzo, T.C. Eschrich, B.N. Engel, and S. Tehrani, 
\bibitem{iswvsmr} S. Urazhdin, N.O. Birge, W.P. Pratt Jr., J. Bass, \Journal{Appl. Phys. Lett.}{84}{1516}{2004}.
\bibitem{berger} L. Berger, \Journal{Phys. Rev.}{B 54}{9353}{1996}.
\bibitem{tsoiprl} M. Tsoi, A.G.M. Jansen, J. Bass, W.C. Chiang, M. Seck, V. Tsoi, and P. Wyder, \Journal{Phys. Rev. Lett.}{80}{4281}{1998}; \textbf{81}, 493(E) (1998).
\bibitem{polianski} M.L. Polianski and P.W. Brouwer, \Journal{Phys. Rev. Lett.}{92}{026602}{2004}.
\bibitem{weber} W. Weber, S. Riensen, and H.C. Siegmann, \Journal{Science}{291}{1015}{2001}.
\bibitem{bazaliyrc} Ya.B. Bazaliy, B.A. Jones, and Shou-Cheng Zhang, \Journal{Phys. Rev.}{B 57}{R3213}{1998}.
\bibitem{fert} D.L. Mills, A. Fert, and I.A. Campbell, \Journal{Phys. Rev.}{B 4}{196}{1971}.
\bibitem{nunez} A.S. Nunez and A.H. MacDonald, cond-mat/0403710 (2004).
\end{thebibliography}
\end{document}